# The measurement problem and the completeness of quantum states

ZUO Zhihong[1]

School of Computer Science and Engineering,

University of Electronic Science and Technology of China

**Abstract:** In this paper, we present a thought experiment that demonstrates that the equivalence of quantum reduced states and statistical mixed states of ensembles is not merely a simple mathematical formulation in quantum mechanics, but rather possesses profound physical underpinnings. Based on the equivalence, we give a possible solution to the quantum measurement problem. We describe quantum measurement process as two-step physical process: one microscopically controllable process which generates an entanglement between the system being measured and a marking system (or property), and one macroscopic process (uncontrollable microscopically) which detects states (or properties) of the marking system. With the solution, we conclude that the measurement postulate is just a corollary of the completeness of quantum states. Our solution is entirely rooted in the traditional formalism of quantum mechanics, requiring no extensions or modifications to it. We also point out that quantum randomness originates from entanglement, and the collapse of the wave function is a subjective process.

## 1. Introduction

The measurement problem is an important issue in researches on the foundation of quantum mechanics [1-15]. It has been studied extensively and seriously, but there are no satisfying answers accepted by most physicists so far [16, 17].

The measurement postulate in quantum mechanics tells us that if a physical system is in a state $\sum_i c_i|a_i\rangle$ (where $\{|a_i\rangle\}$ is a basis of the Hilbert space assigned to the system), then the system will collapse to a state $|a_i\rangle$ with corresponding probability $|c_i|^2$ after being measured in $\{|a_i\rangle\}$ basis. Measurement causes discontinuous change of quantum state, which conflicts with the continuous evolution ruled by Schrodinger equation. Since proposed, it has been subject to extensive questioning and debate, and is considered by some physicists as the crux of the measurement problem [2-4, 7, 10, 18].

[1] Email: zhzuo@uestc.edu.cn

Moreover, there is no further explanation in the measurement postulate for what quantum measurement is and how it occurs. According to von Neumann's interpretation about quantum measurement [1], if a measured system and a measuring device are in the states $\sum_i c_i|a_i\rangle$ and $|m_r\rangle$ respectively before measurement, then the entire system should be in the state $\sum_i c_i|a_i\rangle|m_i\rangle$ (where, the state $|m_i\rangle$ is the state of the measuring device corresponding to the state $|a_i\rangle$ of the measured system) after measurement. That is, the entire system is in a superposition of different states, rather than a specific state within the superposition. Some physicists believe that explaining why the system always randomly presents a specific state with a certain probability after measurement, rather than being in a superposition, is the core of the measurement problem [2, 6, 7, 19, 20,].

Quantum mechanics is believed to be the most fundamental theory of physics. Any observation of the objective world is also a kind of measurement, making the measurement problem a more general issue, namely, why people have never observed the superposition state of macroscopic objects, and why physical systems always appear with certain specific objective properties (such as position, momentum, energy, etc.). Some physicists believe that this broader issue is also a problem that needs to be addressed in the measurement problem [8, 19, 20].

There exist a lot of quests for solutions to the measurement problem. Some solutions attempt to answer these questions within the framework of orthodox quantum mechanics [3, 4, 12, 14, 19-21], while others seek to reinterpret or extend the formalism of quantum mechanics to obtain satisfactory answers [2, 10, 22]. This article attempts to propose a solution to the measurement problem within the framework of orthodox quantum mechanics that the author believes to be relatively satisfactory.

## 2. Two noteworthy aspects in the measurement problem

In quantum measurement, the physical system being measured is a microscopic object (such as a photon or an electron), and the measuring device (such as a photoelectrometer or a fluoroscope) is a macroscopic instrument, therefore, the measurement process as a whole is a macroscopic process.

Quantum mechanics originated from the description of microscopic phenomena, and when it is applied to macroscopic process like measurement, extra caution is required.

Let $A$ be the physical system being measured and $M$ the measuring device; $A$ and $M$ constitute a closed system. If the system $A$ is in a superposition state $\sum_i c_i|a_i\rangle$ and the device $M$ is in an initial state $|m_r\rangle$, according to quantum mechanics, the evolution of the closed system during measurement is governed by the Schrödinger equation. The evolution is determined by a linear operator $U$, which will transform the superposition state of the physical system $A$ being measured into the superposition state of the entire system $A + M$, that is,

$$\left(\sum_i c_i|a_i\rangle\right)|m_r\rangle \xrightarrow{U} \sum_i c_i|a_i\rangle|m_i\rangle.$$

This process, known as process 2 in the von Neumann measurement process, has been explicitly or implicitly adopted in the study of measurement problem since its introduction [1]. However, this derivation is not self-evident and requires further careful analysis.

The measurement process is a macroscopic one, and at the microscopic level, there exist numerous degrees of freedom that are not controlled by the experimenter. These degrees of freedom vary completely during each experimental process. Firstly, the same determined macroscopic state corresponds to many different states at the microscopic level. Each macroscopic state of the measurement device $|m\rangle$ is actually a family of microscopic states $\{|m,\alpha\rangle\}$ ($\alpha$ represents the degrees of freedom at the microscopic level that are not controlled by the experimenter in the macroscopic experiment; for simplicity, it is assumed that these degrees of freedom are discrete). Secondly, the linear evolution ruled by Schrodinger equation also depends on the degrees of freedom uncontrollable by experimenters. Even for a closed system of which the energy remains constant, the particles made up the system and interactions between them are different in every run of the measurement from the microscopic view, resulting in completely different Hamiltonian operators. Therefore, every run of a measurement, while same as a macroscopic process, but it experiences microscopically different linear evolution every time. When the measured system and the measuring device is in a state $(\sum_i c_i|a_i\rangle)|m_r,\alpha'\rangle$, the measurement process causes a linear evolution of the compound system, denoted this evolution operator by $U'$. According to the linearity of $U'$,

$$U'\left(\left(\sum_i c_i|a_i\rangle\right)|m_r,\alpha'\rangle\right) = \sum_i c_i U'(|a_i\rangle|m_r,\alpha'\rangle).$$

Since linear evolution $U'$ is the linear evolution of the measurement when the system is in $(\sum_i c_i|a_i\rangle)|m_r,\alpha'\rangle$, from a microscopic point of view, it is completely different from the linear evolution $U''$ of another measurement when the system is in a specific state $|a_i\rangle|m_r,\alpha''\rangle$. Therefore, $U'(|a_i\rangle|m_r,\alpha'\rangle) = |a_i\rangle|m_i,\alpha'\rangle$ cannot be obtained from $U''(|a_i\rangle|m_r,\alpha''\rangle) = |a_i\rangle|m_i,\alpha''\rangle$ and it cannot be derived that

$$U'\left(\sum_i c_i|a_i\rangle\right)|m_r,\alpha'\rangle = \sum_i c_i|a_i\rangle|m_i,\alpha'\rangle.$$

From the above argument, it is problematic to think that the linear evolution of quantum mechanics will inevitably extend the microscopic superposition state of the system being measured to the macroscopic superposition state of the entire system (the measuring equipment plus the measured system).

Another point needed more attentions pertains to quantum state, especially entangled state. In quantum mechanics, when a physical system is in a definite state$|\varphi\rangle$, for any basis $\{|b_i\rangle\}$ of its associated Hilbert space, $|\varphi\rangle$ can be decomposed as $|\varphi\rangle = \sum_i \lambda_i|b_i\rangle$. This is merely a mathematical expression devoid of physical significance; at this stage, the system does not possess any property (or value) $|b_i\rangle\langle b_i|$ (unless one $|\lambda_i|^2 = 1$ and all others are zero). However, according to the measurement postulate, if a measurement in the basis $\{|b_i\rangle\}$ were to be performed on the system at this moment, it would result in the system being found in one of the states, $|b_i\rangle$ (i.e., possessing the property $|b_i\rangle\langle b_i|$). In other words, a microscopic object in a definite quantum state does not have any properties other than those corresponding to that state; it is the measurement that gives it certain properties that it did not have before the measurement. For composite systems, the situation becomes more complicated. While the composite system is in a determinate state, its subsystems can be in a completely undefined state (i.e., not have any properties at all).

As an example, consider two electrons of which their spins are in a Bell state

$$\Phi = \frac{1}{\sqrt{2}}(|\uparrow\uparrow\rangle + |\downarrow\downarrow\rangle).$$

Here, the spins of both electrons are in an indeterminate state, meaning they have no spin in any direction. This state should not be interpreted as a coexistence of both electrons being in the state $|\uparrow\rangle$ and both being in the state $|\downarrow\rangle$. This is the conclusion of the formalism of quantum mechanics and a fact confirmed by experiments. If the electrons have spin in all directions, the spin values could be considered as hidden variables, leading to the state $\Phi$ satisfying Bell's inequalities [23, 24]. However, all experiments to date confirm that the state $\Phi$ does not satisfy Bell's inequality, in accordance with the predictions of quantum mechanics [25]. It is a corollary of quantum mechanics that subsystems in entangled states do not have definite states (or properties), but it is often ignored, intentionally or unintentionally, when discussing measurements. For example, as Brukner [26] discussed the Wigner's friend problem, he thought definitely that, when electrons, measuring devices and observers is in the state

$$\frac{1}{\sqrt{2}}(|z+\rangle_1|z+\rangle_2|z-\rangle_3|knows\ "up"\rangle_4+|z-\rangle_1|z-\rangle_2|z+\rangle_3|knows\ "down"\rangle_4),$$

the observers are in the definite state $|knows\ "up"\rangle_4$ or definite state $|knows\ "down"\rangle_4$. Frauchiger et al [27] believed improperly that the experimenter in a supposition state is in a certain state (consequently, they have definite ideas and others), and therefore, they obtained incorrectly the conclusion that quantum mechanics cannot describe itself consistently.

The neglection about indefinite state appears also in the decoherence solution to the measurement problem. In the solution, the measured microscopic system, the measuring device and the environment are in the superposition state $\sum_i c_i|a_i\rangle|m_i\rangle|e_i\rangle$. According to quantum mechanics, their states are all indefinite, but this is completely contradiction with our common experiences—the environment presents itself definitely at all time. Moreover, in the decoherence scheme, the measured system and the measuring device is in a mixed state $\sum_i|c_i|^2|a_i\rangle|m_i\rangle\langle a_i|\langle m_i|$ after tracing out states of the environment. The mixed state, however, is from a superposition state in which the states of every system is indefinite. Thus, the mixed state consists of indefinite states, not the expected mixed state that would be composed of definite measurement results.

The measurement problem encompasses various aspects of quantum mechanics, and coupled with the incomplete considerations mentioned above when discussing it, it becomes obscure and complex, which is one of the reasons why it has long eluded a clear and reasonable explanation.

## 3. How to understand quantum measurement

A solution to the measurement problem means that it gives answers to all (or parts) of problems in Introduction, namely, eliminating the measurement postulate in the formalism of quantum mechanics; answering what quantum measurement is, preferably by giving its necessary, if not sufficient, conditions; explaining why macroscopic objects have not been observed in superposition states and what is the origin of the quantum randomness, if macroscopic physical processes are regarded as quantum processes.

The extraordinary success of quantum mechanics is no accident and the seemingly bizarre measurement postulates, confirmed by countless experiments, must inherently possess unparalleled rationality. We believe that quantum mechanics itself contains the essential elements for a reasonable solution of the measurement problem, and what needs to be done is merely to organize and present them.

### *3.1 The completeness of quantum states and the equivalence of reduced states and mixed states*

In history of quantum mechanics, the quantum description of microscopic objects had been strongly suspected by many physicists. Einstein had thought that quantum states are not complete and they do not include all physical elements in objects which they describe. He believed that some requisite 'hidden variables' must be added to quantum states in order to describe microscopic objects completely [28]. By Bell's researches and many experiments, it has been confirmed that 'hidden variable' theories are invalid, and quantum states are complete. Even though not all 'hidden variable' theories of quantum mechanics are eliminated so far, it can be safely said that quantum mechanics is correct about the completeness of quantum state [25, 29].

For a subsystem of a compound system in a definite (pure state), quantum mechanics provides a specific description, reduced state, which can be obtained by tracing out states of other subsystems in the compound system. In general, reduced state is regarded as a convenient tool for calculations of the expected value of observables on subsystems, but not the complete description of them [1, 24, 30].

The key step in our solution to the measurement problem is our insistence on and emphasis of the completeness of quantum states, particularly the completeness of reduced states of subsystems. When a subsystem participates in physical processes as an independent entity, this state encompasses all its properties required for those physical processes.

The idea of reductive state completeness may seem a bit ill-considered at first glance. The immediate intuition stems from the entangled state of spins of two electrons, $\Phi = \frac{1}{\sqrt{2}}(|\uparrow\uparrow\rangle + |\downarrow\downarrow\rangle)$. In terms of quantum mechanics, the reduced states of two electrons are $\rho_1 = \rho_2 = \frac{1}{2}I$, and at this case, their spins are indefinite. But, if an electron (or an ensemble of electrons) is generated in a state $|\uparrow\rangle$ or $|\downarrow\rangle$ each with 50% probability, then it is in a mixed state, the quantum description of which is also $\rho = \frac{1}{2}I$. The two scenarios appear to have obvious differences. In the former situation, the electron's spin is completely indeterminate, while in the latter situation, however, the spin of each electron (in the ensemble) is in a definite state, $|\uparrow\rangle$ or $|\downarrow\rangle$. This seems to suggest that there may be a physical process that can distinguish between the two cases. On the other hand, if quantum states describe completely microscopic objects, then, due to their same quantum states, electrons in two situations will be completely indistinguishable and would have exact same physical effects at all time, namely, there is no any physical process can distinguish between the two cases.

However, we can conceive a communication protocol to show the correctness of the quantum states about the two situations. Suppose that electrons in two cases mentioned above are different on physical effects, that is, there is at least a physical process $P$ which can distinguish them and produce distinct result $r_1$ and $r_2$, respectively. (The process can be implemented for a single electron, or for an ensemble of electrons in a same state.) Suppose that Alice and Bob are far from each other, and they prepare a lot of pairs of electrons in the entangled state $\Phi = \frac{1}{\sqrt{2}}(|\uparrow\uparrow\rangle + |\downarrow\downarrow\rangle)$ before they separate. Alice possesses the electron 1, and Bob possesses the electron 2. They divide pairs of electrons into $n$ groups, each of which has same number of pairs. If Alice intends to send a binary bit 0 to Bob through a group of electron pairs, then she does nothing to the electrons in all pairs she possesses; if she intends to send a binary bit 1 to Bob, then she measures her electrons in the $\{|\uparrow\rangle, |\downarrow\rangle\}$ basis. Bob, in the other end, carries out physical process $P$ on his electrons. For pairs of electrons on which Alice does nothing, Bob's electrons are in the reduced state of $\Phi$, and he gets the result $r_1$ after the process $P$, and thus he gets the bit 0; for pairs of electrons on which Alice measures, according to quantum mechanics electrons collapse into the definite state $|\uparrow\rangle$ or $|\downarrow\rangle$ with 1/2 chance respectively after the measurement, and consequently, the Bob's corresponding electrons are in same state, i.e., these electrons are in a mixed state of $|\uparrow\rangle$ and $|\downarrow\rangle$, and thus belong to the latter case mentioned in the last paragraph. After the process $P$, Bob records the result $r_2$ and gets the bit 1. Thus, this protocol would provide a communication on space-like spacetime distance which is directly in contradiction with special relativity. The correctness of quantum mechanics and special relativity means electrons in two cases are indistinguishable physically. This argument shows convincingly the equivalence of reduced state and mixed state.

Note that the communication protocol depends heavily on the measurement postulate. In the next sub-section, based on the equivalence of reduced states and mixed states we will propose a possible explanation of quantum measurement, and take the measurement postulate as a consequence of the equivalence, and remove it from quantum mechanics. But replacing the measurement postulate with the equivalence is not our best choice. Even though the equivalence is better than the measurement postulate (at least, the equivalence does not refer to any unexplained concepts, such as measurement, measuring device, observer and so on), it is still wired compared with other postulates in quantum mechanics. Fortunately, we do not need to do so. What we do is to insist the completeness of quantum state, and that reduced state and mixed state as special kinds of quantum state, are complete too. Based on it, a reduced state and the same mixed state would have totally same physical effects and are indistinguishable. We do not change anything in quantum mechanics and just emphasize the universal validity of the completeness of quantum state.

### *3.2 Measurement process*

Based on the equivalence of reduced and mixed states, we are able to a give reasonable explanation for quantum measurements.

We illustrate measurement process with an example. Suppose that a 1/2-spin particle has the 'up' spin in $\boldsymbol{a}$ direction (denoted the state as $|\boldsymbol{a}\rangle$), and one experimenter intends to measure its spin in $\mathbf{z}$ axis direction (denoted the 'up' and 'down' spin in $\mathbf{z}$ axis as $|\uparrow\rangle$ and $|\downarrow\rangle$). According to quantum mechanics, $|\boldsymbol{a}\rangle$ can ae represented as $|\boldsymbol{a}\rangle = \alpha|\uparrow\rangle + \beta|\downarrow\rangle$. In this state, however, the particle has not the spin in $\mathbf{z}$ axis (suppose that $\boldsymbol{a} \neq \mathbf{z}$). Physically, an indefinite (precisely, nonexistent) state (or the corresponding property) cannot generate physical effects, that is, it cannot play a role in physical processes and produce definite observable results. Therefore, the first step of quantum measurement has to contain a physical process through which the measured state (or the property) will be marked by a state (or a property) of another particle. In order to make measurement to be possible and precise, the marking would be in one-one correspondence between the original state (or the property) and the marking state (or the property), and the marking state (or property) would be definite and existent. Quantum mechanics provides a very good marking method—entanglement. In our example, we can make the measured particle interacting with another particle, which generates the ready state $|\boldsymbol{a}\rangle_r = \alpha|\uparrow\rangle|0\rangle + \beta|\downarrow\rangle|1\rangle$, in which $|0\rangle$ and $|1\rangle$ are two orthogonal states of the marking particle (for instance, vertical and horizontal polarization, if the marking particle is a photon) which mark $|\uparrow\rangle$ and $|\downarrow\rangle$ states of the measured particle, respectively. Two particles now constitute a compound system. According to quantum mechanics, the reduced state of the marking particle is $|\alpha|^2|0\rangle\langle 0| + |\beta|^2|1\rangle\langle 1|$, and actually it is not in any definite state. Now, here is the crux. According to our argument, the reduces state is equivalent physically to a mixed state, in which the marking particle is in a definite state, $|0\rangle$ or $|1\rangle$ with probabilities $|\alpha|^2$ and $|\beta|^2$ respectively. A definite state (or a property) can participate in physical processes, and can produce physical effects. Thus, the next step of measurement is a macroscopic physical process which takes the particle in definite states (or a property) as input, and produces desired observable results. In our example, the second step is to design a macroscopic physical process in terms of some physical effects, in which, when the marking particle is in state $|0\rangle$, then the pointer of the measuring device would point to the scale 1; when the particle is in state $|1\rangle$, the pointer would point to the scale 2. Because the reduced state of the marking particle is $|\alpha|^2|0\rangle\langle 0| + |\beta|^2|1\rangle\langle 1|$, after this macroscopic process, the pointer would point to the scale 1 with probability $|\alpha|^2$ and the scale 2 with probability $|\beta|^2$. When the marking particle is in the definite state $|0\rangle$ (or $|1\rangle$), the state of the measured particle must be in the corresponding state $|\uparrow\rangle$ (or $|\downarrow\rangle$), that is, if we measure its spin in $\mathbf{z}$ axis once again, we will find the 'up' spin (or 'down' spin) absolutely.

In the above example, in order to describe conveniently, we suppose the entanglement exists between two particles. In actually measurements, the entanglement in most cases occurs between different properties of same particle. For instance, the above 1/2-spin particle in a state $|\boldsymbol{a}\rangle = \alpha|\uparrow\rangle + \beta|\downarrow\rangle$ can be entangled with its locations as $|\boldsymbol{a}\rangle_r = \alpha|\uparrow\rangle|x_1\rangle + \beta|\downarrow\rangle|x_2\rangle$, in which $|x_1\rangle$ and $|x_2\rangle$ are its two locations in space. In quantum mechanics, the description of the composite properties of physical objects is analogous to that of composite systems, where the corresponding Hilbert space is the product of the spaces of each property. The sole requirement is that these properties can be measured simultaneously, i.e., formally, their corresponding Hermitian operators have to be commutable. When considering one property independently, we can obtain its description by tracing out the other properties. Taking as an example, $\mathrm{Tr}_{\mathrm{spin}}(|\boldsymbol{a}\rangle_r) = \mathrm{Tr}_{\mathrm{spin}}(\alpha|\uparrow\rangle|x_1\rangle + \beta|\downarrow\rangle|x_2\rangle) = |\alpha|^2|x_1\rangle\langle x_1| + |\beta|^2|x_2\rangle\langle x_2|$. The argument about the equivalence of reduced state and mixed state can be modified a little to be applied on property, that is, we can design a similarly protocol to finish space-like communications, if reduced property and mixed property are not equivalent.

To sum up, quantum measurement process is a two-step physical process. The first step is a microscope process which entangles the state (or the property) of the measured system with a state (or a property) of a marking system, or with a property of itself. The second step is a macroscopic process which detects a specific state (or a property) of the marking system or itself. It is the reduced state (or the reduced property) that, which is equivalent to a mixed state (or a mixed property), guarantees accordingly that it can take part into physical processes which generate desired results.

### *3.3 Famous experiments in quantum mechanics*

The above interpretation of quantum measurement is universally valid and can be used to explain some famous experiments in quantum mechanics.

In a Stern-Gerlach experiment, an electron (a silver atom in actual experiments) with the 'up' spin in $\mathbf{y}$ axis flies along $\mathbf{x}$ axis into an inhomogeneous magnetic field in the $\mathbf{z}$ axis. By this physical process, the 'up' spin in $\mathbf{y}$ axis which can be resolved into spins in $\mathbf{z}$ axis as $|\mathbf{y}_+\rangle = \frac{1}{\sqrt{2}}(|\mathbf{z}_+\rangle + i|\mathbf{z}_-\rangle)$ causes the entanglement between its spins in $\mathbf{z}$ axis and its paths through the magnetic field, resulting in the entangled state $|\mathbf{y}_+\rangle_r =$ $\frac{1}{\sqrt{2}}(|\mathbf{z}_+\rangle|upper\rangle + i|\mathbf{z}_-\rangle|lower\rangle)$, where $|upper\rangle$ and $|lower\rangle$ represent two paths through which the electron passes. The detector screen is used to check which path the electron passes through but not the electron's spin, so what do works is the path information, that is, the reduced property $\mathrm{Tr_{spin}}(|\mathbf{y}_+\rangle_r) = \frac{1}{2}|upper\rangle\langle upper| + \frac{1}{2}|lower\rangle\langle lower|$. Thus, the detector screen finds that half electrons appear in the upper path and the other half in the lower path. In conclusion, when measured the $\mathbf{z}$ axis spin of electrons with the 'up' spin in $\mathbf{y}$ axis, half electrons have 'up' spin in $\mathbf{z}$ axis, and half electrons have 'down' spin in $\mathbf{z}$ axis.

In a Mach-Zehnder interference experiment, after a photon goes through the first half-silvered mirror, it propagates along two paths, upper and lower, to the photon receivers. In orthodox quantum mechanics, the photon's state is described by $|photon\rangle = \frac{1}{\sqrt{2}}(|upper\rangle + |lower\rangle)$, which is a superposition state and is used to explain the interference when the second silvered mirror is inserted. Here, the two photon receivers finish the detection of which path the photon pass through.

However, photon receivers only check the photon's position where the photon arrives at, but not the paths through which it goes. It is the correspondence between the position and its path that points which path the photon passes. So, more completely, in the course of the photon's propagating from the first half-silvered mirror to receivers (or the second half-silvered mirror), the photon is in a state $|photon\rangle = \frac{1}{\sqrt{2}}(|upper\rangle|x_1(t)\rangle + |lower\rangle|x_2(t)\rangle)$ at every moment $t$, in which $|x_1(t)\rangle$ and $|x_2(t)\rangle$ are positions of the photon on the corresponding path at the moment $t$. When there is no second half-silvered mirror, photon receivers detect the photon's position, which can be obtained by tracing out the paths, i.e., $\frac{1}{2}|x_{receiver1}\rangle\langle x_{receiver1}| + \frac{1}{2}|x_{receiver2}\rangle\langle x_{receiver2}|$, where $x_{receiver1}$ and $x_{receiver2}$ are the positions of two photon receivers, respectively . So, the photon has half chance to be at $x_{receiver1}$ and half chance at $x_{receiver2}$, that is, the photon goes along two paths each with 50% probability.

On the other hand, if the second half-silvered mirror is inserted, the photon will arrives at the same position on the mirror, and its state now is $|photon\rangle = \frac{1}{\sqrt{2}}(|upper\rangle + |lower\rangle)|x_{second-mirror}\rangle$ in which the photon's position state $|x_{second-mirror}\rangle$ is separated out and the photon is in a path superposition state. Now, the second mirror causes the interference.

The Born interpretation of wave function can be explained in similar way. Take the two-slit experiment as an example. When a photon (or an electron) passes along $\boldsymbol{y}$ axis through two slits, it is in a superposition state $|photon\rangle = \frac{1}{\sqrt{2}}(|slit_1\rangle + |slit_2\rangle)$, in which $|slit_1\rangle$ and $|slit_2\rangle$ represents the state in which the photon takes across the slit 1 or slit 2, respectively. The receiving screen is placed along $\boldsymbol{x}$ axis. For simplicity, suppose that location states $\{|x\rangle\}$ form a complete basis, i.e., $\int |x\rangle\langle x| dx = I$. The photon's state can be resolved along $\boldsymbol{x}$ axis as $|photon\rangle = \int \big(w_1(x) + w_2(x)\big)|x\rangle dx$, in which $w_1(x) = \frac{1}{\sqrt{2}}\langle slit_1|x\rangle$ and $w_2(x) = \frac{1}{\sqrt{2}}\langle slit_2|x\rangle$ are wave functions corresponding to slit 1 and slit 2, respectively. While the photon moves to the receiving screen from slits, experiments can detect the location in $\boldsymbol{x}$ axis at each moment. Thus, the photon has path property, even though in a superposition of paths, the path is indefinite. Therefore, the complete description of the photon in the course of moving from slits to the receiving screen should include the path information, that is, the photon's state is $|photon\rangle = \int \big(w_1(x) + w_2(x)\big)|path_x\rangle|x\rangle dx$, in which $|path_x\rangle$ is the path through which the photon passes from slits to the $x$ position on the screen. It is an entangled state. The screen only detects the photon's location, but not paths, so tracing out path states, the state of the photon's location is $\int |w_1(x) + w_2(x)|^2 |x\rangle\langle x| dx$. So, the probability of receiving the photon at some $\boldsymbol{x}$ interval on the screen is $|w_1(x) + w_2(x)|^2 dx$. It is the Born interpretation of wave function.

The explanation of measurement process in fact establishes a necessary condition for quantum measurement. According to the condition, some common devices in quantum experiments are not measuring devices, such as half-silvered mirror in Mach-Zehnder interferometer and magnet devices in Stern-Gerlach experiments, because they just finish the first step of quantum measurement, but lack the second step. It coincides with our practical experiences in quantum experiments.

## 4. Randomness, definite unitary evolution, collapse of wave function and superposition state of macroscopic object

Randomness is essential in quantum mechanics and manifests in the measurement postulate, but there is no any explanation for its origin in quantum mechanics.

However, it is known from the exposition of quantum measurements in the last section that, even though quantum objects evolve according to definite Schrodinger equation and definite states cannot evolve into indefinite states, but due to the equivalence of reduced state and mixed state, the statistical feature in mixed state leads to the occurrence of randomness in quantum mechanics when states (or the microscopic systems with the states or properties) in mixed state participate in physical process. Because entangled states between subsystems generate truly mixed states (for a compound system in a product state, its subsystem is in a pure state, not a mixed state), it is entanglement which causes randomness, and it is the equivalence of reduced state and mixed state which brings randomness into reality.

If the system being measured and the measuring devices are collectively regarded as a closed system, then quantum measurement process can be considered as a combination of two deterministic unitary evolutions. The first unitary evolution generates an entangled state of the measured system and the 'marking' system (or propertiy of the measured system). The second unitary evolution is the macroscopic physical process which finishes the interaction between the 'marking' system and the measuring device. The first process is controllable microscopically, but the second process is not. As we stressed on in section 2, the latter macroscopic unitary evolutions are microscopically different in its every run. This kind of unitary evolutions cannot cause the superposition of macroscopic states. Moreover, despite the randomness in measurement outcomes, the overall evolution of the entire system is definite and unitary, and thus is not in contradiction with the definite dynamics of quantum mechanics.

If the measured system is in a state $\sum_i c_i|a_i\rangle$, according to our interpretation of quantum measurement, after the first step, it entangles with the 'marking' system and its state is $\sum_i |c_i|^2 |a_i\rangle\langle a_i|$ if it is considered independently as a single system. After the second macroscopic process, the 'marking' system is determined to be a state $|m_i\rangle$, then the measured system is in the corresponding state $|a_i\rangle$. As an independent single system, it's state varies in a whole measurement as

$$\sum_i c_i|a_i\rangle \rightarrow \sum_i |c_i|^2 |a_i\rangle\langle a_i| \rightarrow |a_i\rangle.$$

This state's change is called collapse of wave function in quantum mechanics. From our interpretation of quantum measurement, the collapse occurs in the second step of quantum measurement and it is only a change in observer's knowledge, that is, if we adopt the ignorance interpretation of mixed state, then the state's change is caused by the change of observer's knowledge about system's state and does not happen physically, but epistemologically. This explain why the collapse never be observed in experiments. From physical view, the state of the measured system is always $\sum_i |c_i|^2 |a_i\rangle\langle a_i|$. Only for the observe who knows that the state of the 'marking' system is $|m_i\rangle$, the state of the measured system is $|a_i\rangle$. Therefore, there is no collapse of wave function, but variation of observer's knowledge.

No matter in the interpretation of quantum measurement in orthodox quantum mechanics or in our interpretation, macroscopic physical processes cannot generate superpositions of macroscopic states. In order to expanse a superposition of microscopic states to a superpositions of macroscopic states, a unitary evolution $U$ must satisfy $U(\sum_i c_i |a_i\rangle)|m_r, \alpha\rangle = \sum_i c_i U(|a_i\rangle |m_r, \alpha\rangle) = \sum_i c_i |a_i\rangle |m_i, \alpha\rangle$. It must be the same evolution for all different states $|a_i\rangle$ of the measured system and all initial states $|m_r, \alpha\rangle$ of the measuring device. Moreover, for the unitarity, it is required essentially to be defined on all states of the measured system and the measuring device. This requirement necessitates $U$ to control all degrees of freedom in the measuring device. Due to tremendous numbers of degrees of freedom in a macroscopic device, designing this kind of unitary operators, if not impossible, is extremely difficulty. The superposition state of macroscopic object exists only in theories or thought experiments so far.

## 5. Conclusions

We give a possibly promising solution to the quantum measurement problem. It is encompassed within orthodox quantum mechanics, without any extension or modification to the formalism of quantum mechanics, nor does it provide any new or strange interpretations or quantum mechanics.

The crux of our solution lies in emphasizing that quantum states, as formalized means of describing physical objects, are complete. Specifically, when a subsystem of a composite system is treated as an independent physical system, its reduced state is also complete, leading to its equivalence with statistical mixed states, which we refer to as the equivalence of quantum states. The argument for the state equivalence is built upon two highly successful physical theories, quantum mechanics and special relativity, and we consider the argument to be compelling.

Based on the equivalence of reduces state and mixed state, we elaborate on quantum measurement and take the measurement postulate as its corollary. We describe measurement process as two-step physical process: one microscopically controllable process which produces entanglement, and one macroscopic process (uncontrollable microscopically) which detects given states (or properties) of some microscopic objects. It constitutes the necessary condition of quantum measurement, and thus give an answer to what quantum measurement is.

We also point out that quantum randomness comes from reduced state superficially and in fac from entanglement deeply. As a compound system is in a pure state, the reduced states of its subsystems are equivalent to some mixed states which are probabilistic. It is the probabilistic feature of mixed state which embodies quantum randomness. Moreover, the quantum state of a measured system is a reduced state after measurement, and duo to the equivalence of reduced state and mixed stat, when an observer knows the measurement result, the measured system is regarded as in a definite state. Thus, the collapse of wave function is not an objective physical process, but a subjective change of observer's knowledge.

We take the completeness of quantum states as a fundamental principle, treating state equivalence, i.e., the equivalence of reduced and mixed states, as a conclusion derived from it. However, state equivalence itself is well worth investigating. Our argument only demonstrates the necessity of this equivalence, but it does not shed light on why it is the case. In quantum mechanics, quantum state is the most fundamental, with no deeper concept beneath it. Therefore, it seems impossible to explain the equivalence in quantum mechanics. This situation is similar to the equivalence of inertial reference frames in special relativity and the equivalence of non-inertial reference frame and the gravity field in general relativity, they as the basic principles in two theories respectively, cannot be explained in the theories based on them. However, as Bell's stated in his discussion on quantum nonlocality, the scientific attitude is 'crying out for explanation'. For the equivalence of reduced and mixed states, we anticipate deeper research to achieve a more profound understanding and explanation about it.

**Acknowledgments:** The author would like to thank all members in Lab of Computing Intelligence in SCSE of UESTC for their supporting to this work.